\documentclass[floatfix,nofootinbib, secnumarabic,amssymb,bibnotes, aps,prl,reprint,amsmath,superscriptaddress]{revtex4-2}
\usepackage{epsfig}
\usepackage{graphicx}
\usepackage{dcolumn}
\usepackage{bm}
\usepackage{CJK}
\usepackage{float}
\usepackage{tabularx}
\usepackage{hyperref}
\usepackage{threeparttable}
\usepackage{multirow}
\usepackage{xcolor} 
\usepackage[T1]{fontenc} 
\newcommand{\header}[1]{{\em#1.---}}	

\usepackage[normalem]{ulem}

\begin{document}
\begin{CJK}{GBK}{song}
\title{Direct observation of three-neutron emission from $^7$He$^*$ and the search for the trineutron}

\def \pku      {School of Physics and State Key Laboratory of Nuclear Physics and Technology, Peking University, Beijing 100871, China}
\def \rnc      {RIKEN Nishina Center, Hirosawa 2-1, Wako, Saitama 351-0198, Japan}
\def \lpc      {Universit\'e de Caen Normandie, CNRS/IN2P3, ENSICAEN, LPC Caen, UMR6534, 14050 Caen Cedex, France}
\def \tudarm {Institut f\"ur Kernphysik, Technische Universit\"at Darmstadt, D-64289 Darmstadt, Germany}
\def \gsi      {GSI Helmholtzzentrum f\"ur Schwerionenforschung, 64291 Darmstadt, Germany}
\def \idpno  {Institut de Physique Nucl\'eaire Orsay, IN2P3-CNRS, 91406 Orsay Cedex, France}
\def \tum     {Technische Universit\"at M\"unchen, 85748 Garching, Germany}
\def \dddp   {Departamento de F\'isica de Part\'iculas and IGFAE, Universidade de Santiago de Compostela, E-15782 Santiago de Compostela, Spain}
\def \hku      {Department of Physics, The University of Hong Kong, Hong Kong, China}
\def \kvi       {ESRIG, University of Groningen, Zernikelaan 25, 9747 AA Groningen, The Netherlands}
\def \mta      {Institute for Nuclear Research (HUN-REN ATOMKI), Debrecen, Hungary}
\def \uoy      {Department of Physics, University of York, York YO10 5DD, United Kingdom}
\def \lund     {Department of Physics, Lund University, 22100 Lund, Sweden}
\def \rikkyo   {Department of Physics, Rikkyo University, 3-34-1, Nishi-Ikebukuro, Toshima, Tokyo 171-8501, Japan}
\def \rbsi       {Ru\dj er Bo\v skovi\'c Institut (RBI), Zagreb, Croatia}
\def \cea       {D\'epartement de Physique Nucl\'eaire, IRFU, CEA, Universit\'e Paris-Saclay, 91191 Gif-sur-Yvette, France}\textit{}
\def \toho      {Toho University, Tokyo 143-8540, Japan} 
\def \titec       {Department of Physics, Institute of Science Tokyo, 2-12-1 O-Okayama, Meguro, Tokyo 152-8551, Japan}
\def \ehwa     {Department of Physics, Ehwa Womans University, Seoul 03760, Republic of Korea}
\def \miya      {Faculty of Engineering, University of Miyazaki, Miyazaki 889-2192, Japan}
\def \tohoku   {Department of Physics, Tohoku University, Aramaki Aza-Aoba 6-3, Aoba, Sendai, Miyagi 980-8578, Japan}
\def \cns        {Center for Nuclear Study, University of Tokyo, 2-1 Hirosawa, Wako, Saitama 351-0198, Japan} 
\def \ifk          {Institut f\"ur Kernphysik, Universit\"at zu K\"oln, K\"oln, Germany}
\def \sait        {Department of Physics, Saitama University, Shimo-Okubo 255, Sakura-ku, Saitama-shi 338-8570, Japan}
\def \ioss       {Institute of Space Sciences, Magurele, Romania}
\def \ganil      {Grand Acc\'el\'erateur National d'Ions Lourds (GANIL), CEA/DRF-CNRS/IN2P3, Bvd Henri Becquerel, 14076 Caen, France}
\def \kyoto     {Department of Physics, Kyoto University, Kitashirakawa, Sakyo, Kyoto 606-8502, Japan}
\def \uod   {Institute of Physics, Faculty of Science and Technology, University of Debrecen, Egyetem t\'er 1., H-4032 Debrecen, Hungary}
\def \ibs     {CENS, Institute for Basic Science, Daejeon, 34126, Republic of Korea}
\def \fudanu   {Key Laboratory of Nuclear Physics and Ion-Beam Application (MOE), Institute of Modern Physics, Fudan University, Shanghai 200433, China}
\def \impa   {Institute of Modern Physics, Chinese Academy of Sciences, Lanzhou 730000, China}
\def \impb   {School of Nuclear Science and Technology, University of Chinese Academy of Sciences, Beijing 100049, China}
\def \impc   {Southern Center for Nuclear-Science Theory (SCNT), Institute of Modern Physics, Chinese Academy of Sciences, Huizhou 516000, China}

\author{S.~W.~Huang}\affiliation{\pku}\affiliation{\rnc} 
\author{C.~Lenain}\affiliation{\lpc}
\author{Z.~H.~Yang}\email{Corresponding author: zaihong.yang@pku.edu.cn}\affiliation{\pku}\affiliation{\rnc} 
\author{F.~M.~Marqu\'es}\affiliation{\lpc}
\author{J.~Gibelin}\affiliation{\lpc}
\author{J.~G.~Li}\affiliation{\impa}\affiliation{\impb}\affiliation{\impc}
\author{A.~Matta}\affiliation{\lpc}
\author{N.~A.~Orr}\affiliation{\lpc}
\author{N.~L.~Achouri}\affiliation{\lpc}
\author{D.~S.~Ahn}\affiliation{\rnc} \affiliation{\ibs}
% ...

\date{\today}
\begin{abstract}

Three-neutron emission from $^7$He has been directly measured for the first time, following neutron knockout from a $^8$He beam at 156~MeV/nucleon. A resonance-like structure at $2.08(4)$~MeV above the $^4$He+$3n$ threshold [$E_x=2.68(4)$~MeV] with a width of $3.9(2)$~MeV was observed and deduced to arise predominately from the predicted $J^{\pi}=3/2^{-}_2$ level.
The three-neutron invariant-mass spectrum was reconstructed and found to peak at around 1~MeV and could, through complete simulations incorporating neutron-neutron correlations, be very well described by the sequential decay of $^7$He$^*$ via the $2_1^+$ excited state of $^6$He. No evidence was found for any significant three-neutron correlations beyond those expected from well-established two-body interactions, including a trineutron resonance.

\end{abstract}

\maketitle
\end{CJK}

\header{Introduction} 
Light nuclei with extreme neutron-to-proton asymmetry provide an ideal testing ground for our understanding of nuclei, including, in particular the details of the two- and few-body forces used in models \cite{Ye25}. In this context, the search for the most extreme systems, such as multineutron clusters or resonances, has been the object in recent years, beginning with the measurement of Refs.~\cite{Mar02,LPCarXiv}, of significant experimental and theoretical efforts \cite{Marques21,Huang23} and culminated with the recent observation of a near threshold resonance-like structure in the $4n$ system \cite{Kisa16, Duer22}.  It is, however, not yet clear whether this structure corresponds to a 4-body resonance \cite{Pieper03,Shirokov16,Gandolfi17,Fossez17,Li19} or is the result of initial-state correlations in the projectile \cite{Lazauskas23} and/or the reaction mechanism \cite{Muzalevskii25}.
For the $3n$ system, the existence of a trineutron resonance has been predicted by $ab$ $initio$ calculations \cite{Gandolfi17, Li19, Mazur24} but no evidence for any such relatively narrow structure has been found in high-statistics missing-mass experiments \cite{Miller80, Miki24}.
An important complementary means to search for multineutron systems is the direct detection of multiple neutron emission from well-defined states.
Besides being independent of the projectile/initial state and the reaction, such direct detection has the potential to enable the investigation of multineutron correlations.

Significant progress has been made in investigating experimentally $2n$ and $2p$ emissions \cite{Zhou22,Kohley12, Spyrou12, Kohley13, Kondo16, Mont24, Kryger95, Mukha07, Egorova12, Wamers14, Brown14, Webb19,Pfutzner23}. On the proton-rich side, several $3p$ emitters have been identified and studied \cite{Char11, Char21, Brow17, Kost19, Xu25} and two $4p$ emitters have been found \cite{Charity10, Jin23}.
In contrast, multineutron emission remains largely unexplored, owing primarily to the limited capabilities for detecting more than two neutrons.  
Recent advances in large-scale neutron arrays have enabled the direct detection of $3n$ and $4n$ emission from $^{27,28}$O \cite{Kondo23}, both decaying sequentially through $^{26}$O \footnote{Evidence has also been found for sequential $3n$ decay of $^{25}$O \cite{Sword19} and $^{15}$Be \cite{Kuchera24}.}. 

Neutron-rich helium isotopes, composed of an $\alpha$ core plus several valence neutrons, provide an ideal laboratory for exploring multi-neutron correlations.
The ground state (g.s.) of $^{6}$He exhibits pronounced dineutron clustering \cite{Sun22, Papadimitriou11}, and a strongly correlated four-neutron subsystem has been suggested to be present in the g.s. of $^{8}$He in a recent $^{8}$He$(p,p\alpha)$ experiment \cite{Duer22}.
Positioned between them, $^{7}$He is thus a promising candidate for probing $3n$ correlations. 
While its g.s. is well established as a low-lying $^{6}$He+$n$ resonance with spin parity $J^{\pi}=3/2^-$ \cite{nndc,Aksyutina09, Cao12, Renzi16}, the excited states and their decay properties remain the subject of considerable debate \cite{Kor99, Kor00, Bohlen01,Wuosmaa05,Rogachev04, Boutachkov05,Beck07,Wuosmaa08,Aksyutina09,Cao12,Meister02,Skaza06,Rotureau06,Fossez18,Jaganathen17,Renzi16,Pieper04,Rodkin23, Mazur22,Baroni13,Baroni13b,Volya05,Myo07,Myo09,Wurzer98}.
Among them, a $3/2^-$ level above the $3n$-emission threshold is of particular interest, since its $J^{\pi}$  matches that favored in predictions for a possible trineutron resonance \cite{Gandolfi17, Li19, Mazur24, Lazauskas05, Hiyama16, Ishikawa20, Higgins20}, making trineutron emission a possible decay mode.
A sequence of excited states ($1/2^-$, $5/2^-$, $3/2^-$) has been consistently predicted by a wide range of models, from the phenomenological shell model \cite{Wuosmaa05} to more sophisticated models including those incorporating the continuum \cite{Pieper04, Rodkin23, Mazur22, Baroni13, Baroni13b, Volya05, Myo07, Myo09}.
Experimentally, however, evidence for these levels remains incomplete and sometimes contradictory.
A narrow $1/2^-$ resonance at $E_x\sim0.6$~MeV has been reported in high-energy breakup and $(p,d)$ reactions on $^8$He \cite{Meister02, Skaza06},  but not observed in many other experiments \cite{Kor99, Kor00, Bohlen01, Wuosmaa05, Rogachev04, Boutachkov05, Beck07, Wuosmaa08, Aksyutina09, Cao12, Renzi16}. Broader structures at $E_x\approx2-3$ MeV have been suggested from neutron removal and isobaric-analog studies, with tentative $1/2^-$  or $5/2^-$  assignments \cite{Rogachev04,Boutachkov05,Wuosmaa05,Renzi16,Wuosmaa08,Kor99,Kor00}. Notably, none of these experiments detected all three neutrons, leaving the decay mechanism and possible signatures of $3n$ correlations unresolved.

In this Letter, we report on the first direct measurement of $3n$ emission from $^7$He excited states populated using the $^8$He$(p,pn)$ reaction. High statistics and excellent invariant-mass resolution were achieved, by combining an intense $^8$He beam, a thick liquid hydrogen target incorporating reaction vertex reconstruction and a state-of-the-art neutron array.  A resonance-like structure decaying exclusively into $^4$He+$3n$ was observed and was identified as arising predominately from the $3/2^-_2$ level based on the momentum distribution of the knocked-out neutron and structure model predictions.
The reconstructed $3n$ invariant-mass spectrum, as well as that for $^4$He+$2n$, demonstrated that the $3n$ emission proceeds sequentially via $^6$He$^*(2_1^+)$. No evidence was found for any significant three-neutron correlations, including a trineutron resonance.

\begin{figure}[htp]
\includegraphics[width=0.99\linewidth]{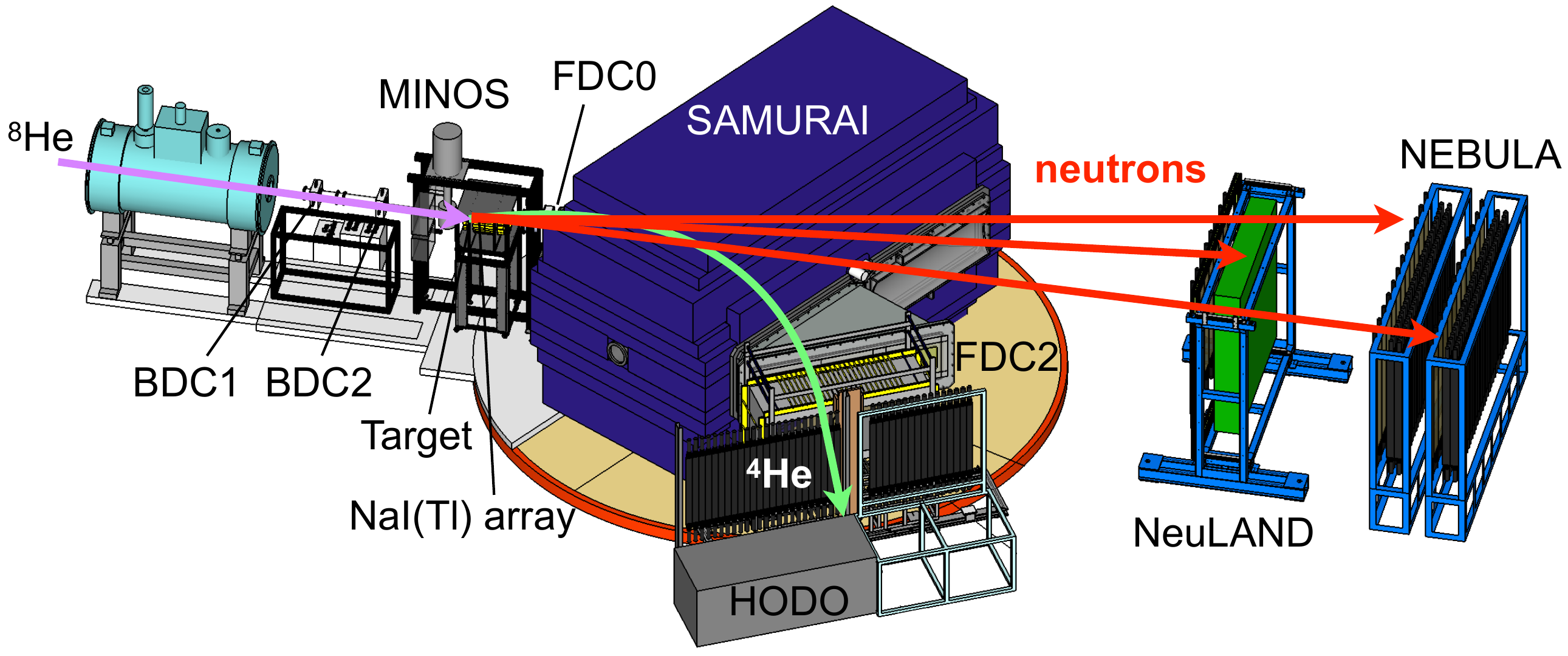}
\caption{Schematic view of the experimental setup. The incident $^8\mathrm{He}$ beam was tracked by two drift chambers (BDC1, BDC2) onto the liquid hydrogen target system (MINOS). Recoil protons from the $(p,pn)$ reaction were tracked using a cylindrical time projection chamber surrounding the target, and their energies were measured by an array of 36 NaI(Tl) crystals. The charged fragments $^4\mathrm{He}$ ($^6\mathrm{He}$) were momentum analyzed using the SAMURAI spectrometer and two drift chambers (FDC0, FDC2), and detected in a plastic scintillator array (HODO). The forward-going beam-velocity neutrons were detected by the multi-element NeuLAND (single-wall) and NEBULA (two separated walls) plastic scintillator arrays, located approximately 11 m and 14 m downstream of the target, respectively.}
\label{fig:Setup}
\end{figure}

\header{Experiment} 
The experiment was performed at the RI Beam Factory operated by the RIKEN Nishina Center for Accelerator-Based Science and CNS, the University of Tokyo. The $^8$He secondary beam ($156$~MeV/nucleon, $\sim$$10^{5}$~pps) was produced by fragmentation of an intense 220~MeV/nucleon $^{18}$O beam on a Be target and purified using the BigRIPS fragment separator \cite{Kubo03, Ohn10}. A schematic view and description of the setup can be found in Fig.~\ref{fig:Setup}. The secondary beam particles were tracked onto the 15~cm thick liquid hydrogen target of MINOS \cite{Ober14} using two drift chambers. The recoil protons from the $(p,pn)$ reaction were tracked using a cylindrical time projection chamber surrounding the target, covering approximately 20--60 degrees. Combined with the trajectory of the incident $^{8}$He ions, this allowed for the reconstruction of the reaction vertex \cite{Sant18}.

The charged fragments $^{4,6}\mathrm{He}$ were momentum analyzed using the SAMURAI spectrometer \cite{Koba13,Sato13}. The momenta of forward-going neutrons were determined from their time-of-flight ($\sigma \approx 160$ ps) and hit positions measured by the NeuLAND \cite{Bore21} and NEBULA \cite{Naka16} plastic scintillator arrays. The combination of these two arrays (with a total thickness of 88~cm) improved significantly the multineutron detection efficiency, enabling the direct detection of three beam-velocity neutrons.
Crosstalks---neutrons scattered within and detected more than once in the neutron arrays---were eliminated in the offline analysis using a rejection algorithm specifically developed for the present three-wall configuration \cite{Kond20, Huang21}. 

The relative energies of the $^{6}$He+$n$, $^{4}$He+$2n$ and $^{4}$He+$3n$ decay channels were reconstructed from the momenta of the He fragments and the coincident decay neutrons using the invariant-mass method.
The $1n$, $2n$, and $3n$ efficiencies varied smoothly with the relative energy (Fig.~S1~\cite{supp}), reaching some 40\%, 14\%, and 3\%, respectively, at 1~MeV. The relative-energy spectra were fit using Breit-Wigner (BW) line shapes with
energy-dependent widths for the resonances and a non-resonant contribution (arising from the non-resonant continuum and, for the $^{4}$He+$2n$ channel, uncorrelated neutrons from the decay of $^7$He to $^6$He$^*$) estimated using a simulation of the non-interacting $^{A}$He+$xn$ system \cite{Mont19}.  
All lineshapes were input into a GEANT4 simulation of the complete setup, which incorporated the geometrical acceptances, efficiencies, and resolutions of the individual detectors, as well as the characteristics of the $^{8}$He beam, target and reaction effects. The simulated data were then analyzed in the same manner as the experimental data, including the crosstalk rejection.  As such, the residual crosstalk contamination was included in the simulated spectra and was estimated to be $\sim$2\% for the $2n$ channel and $\sim$7\% for the $3n$ channel.  The uncertainties reported below include both statistical and systematic components, with the latter evaluated by varying the fit range, parameters of the crosstalk rejection algorithm, and potential fragment-neutron momentum misalignments.

\begin{figure}[htp]
\includegraphics[width=0.99\linewidth]{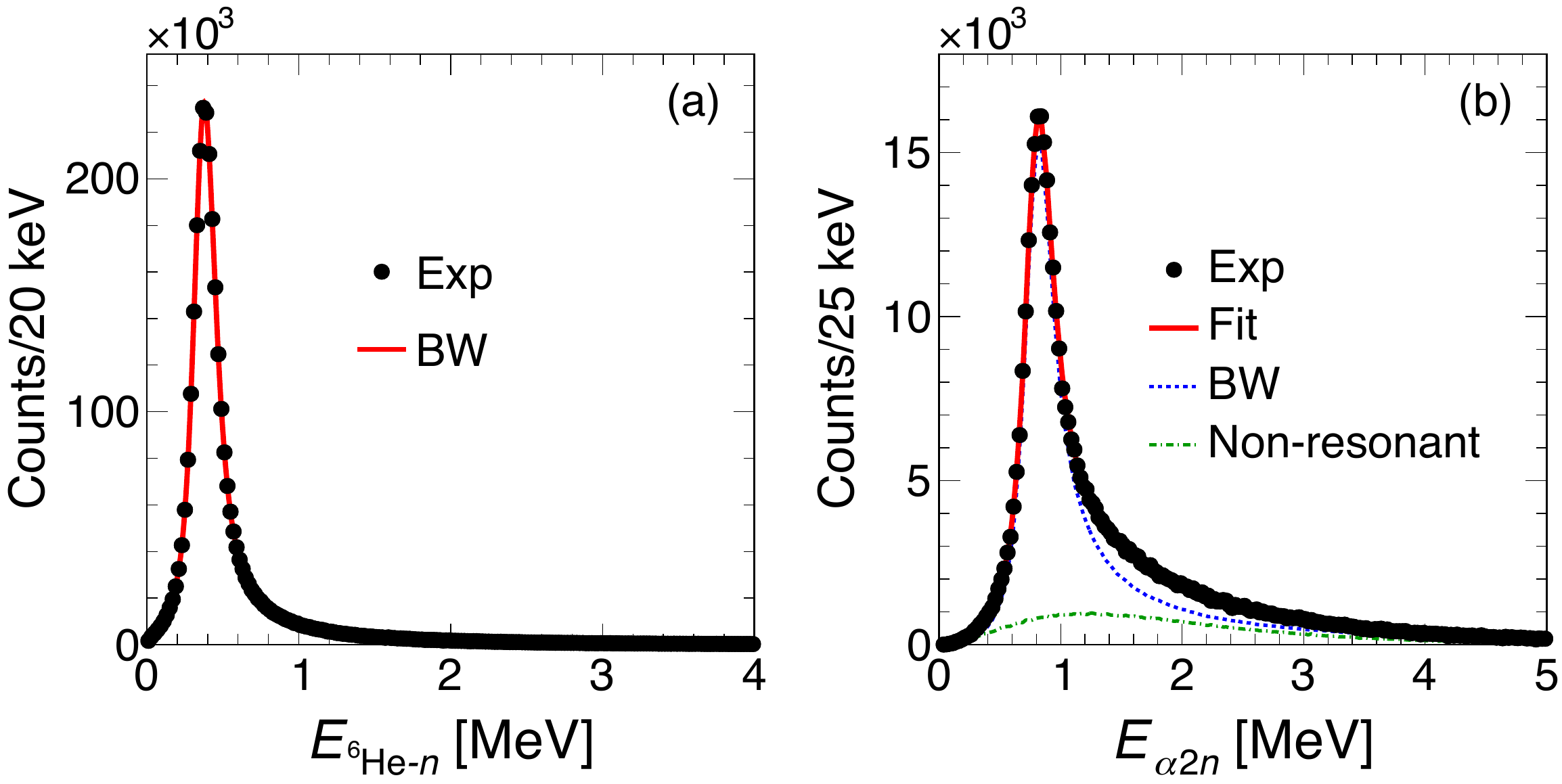}
\caption{Relative-energy spectra for (a) $^6$He+$n$ and (b) $^4$He+$2n$ events. The red line shows the best fit (see text). The vertical error bars are statistical only.}
\label{fig:1n2n}
\end{figure}

\header{One- and two-neutron channels} 
The $^{6}$He+$n$ relative-energy spectrum (Fig.~\ref{fig:1n2n}a) exhibits a single narrow structure near threshold, corresponding to the well-known $^7$He g.s. \cite{nndc}.  
The resolution (FWHM),  determined from simulations,  varies as $\sim$0.16$E_{\rm{^6He}-\it{n}}^{0.57}$~MeV, and is 110~keV at $E_{\rm{^6He}-\it{n}}$=0.5~MeV. The spectrum is well described by a single $\ell$=1 BW resonance  with an energy of $E_r$=379(2)~keV and a width of $\Gamma_r$=137(10)~keV, in agreement with previous measurements of proton-induced neutron knockout from $^{8}$He, albeit with a narrower width \cite{Aksyutina09,Cao12}. The high statistics and broad acceptance of this measurement have allowed us to rule out the population of any higher-lying resonances in this decay channel with a bin-by-bin significance well below $1\sigma$ (details in \cite{supp}).

The $\alpha+2n$ spectrum (Fig.~\ref{fig:1n2n}b) exhibits a narrow peak at $\sim$0.8~MeV above threshold, corresponding to the well-known $2^+_1$ state of $^6$He \cite{nndc}. The resolution varies as $\sim$$0.17E_{\alpha2n}^{0.56}$~MeV and is 150~keV at $E_{\alpha2n}$=0.8~MeV. 
The spectrum was fit up to 3~MeV using a combination of a BW resonance\footnote{$\frac{d \sigma}{d E_{\alpha2n}} \propto \frac{\Gamma(E_{\alpha2n})}{(E_{r}-E_{\alpha2n})^{2}+[\Gamma(E_{\alpha2n}) / 2]^{2}}$, with $\Gamma(E_{\alpha2n}) = \Gamma_{r} \frac{E_{\alpha2n}^{2}}{E^{2}_{r}}$.} and the non-resonant component described above. The red solid line shows the best fit, with $E_r$=823(6)~keV and $\Gamma_r$=228(29)~keV, in agreement with the known energy but with a broader width. 

\begin{figure}[htp]
\includegraphics[width=0.95\linewidth]{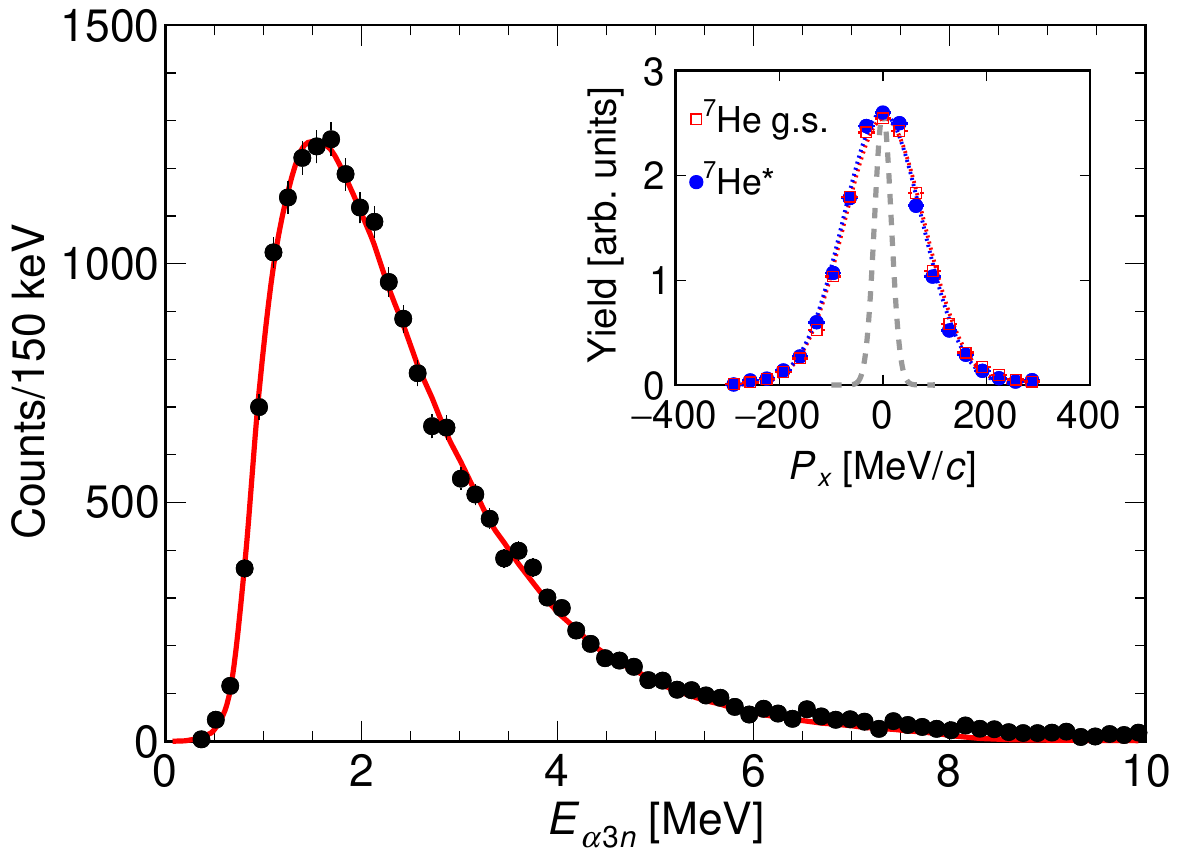}
\caption{Relative-energy ($E_{\alpha3n}$) spectrum for $^4$He+$3n$ events. The red solid line shows the best fit up to 5~MeV using a BW lineshape (see text). The inset shows the transverse momentum distribution ($P_X$) of the knocked-out neutron for $^7$He g.s. ($E_{^6\rm {He}}$$_{-n}<1$~MeV in Fig.~\ref{fig:1n2n}a) and for the present $^7$He$^*$ structure ($1<E_{\alpha3n}<5$~MeV) normalized to the total number of events, with Gaussian fits to the central region to guide the eye. The resolution in $P_X$ (FWHM = 38 ~MeV/$c$) is also shown for comparison (dashed line). The vertical error bars are statistical only.}
\label{fig:3n}
\end{figure}

\header{Three-neutron channel} 
The $^{4}$He+3$n$ relative-energy spectrum (Fig.~\ref{fig:3n}) exhibits a broad structure peaking at $\sim$1.5~MeV. The resolution varies as $\sim$$0.19E_{\alpha3n}^{0.6}$~MeV.
The spectrum was fit up to 5~MeV using a BW resonance\footnote{$\frac{d \sigma}{d E_{\alpha3n}} \propto \frac{\Gamma(E_{\alpha3n})}{(E_{r}-E_{\alpha3n})^{2}+[\Gamma(E_{\alpha3n}) / 2]^{2}}$, with $\Gamma(E_{\alpha3n}) = \Gamma_{r} \frac{\sqrt{E_{\alpha3n}} }{ \sqrt {E_{r} }} $.}, resulting in best-fit parameters of $E_r$=2.08(4)~MeV and $\Gamma_r$=3.9(2)~MeV. 
The simulations employed to produce the lineshapes for the fitting incorporated the sequential decay process that characterizes this structure (see below).
We note that any non-resonant contribution was found to be negligible.  The energy and width found here are in line with, but considerably more precise than, those found in earlier $p$($^8$He, $d$) experiments \cite{Kor99,Skaza06}. 
By comparing the yields of the $^6$He+$n$ and $^4$He+$3n$ channels in the energy range of the present structure, we estimate any potential branching ratio to the $^6$He+$n$ channel to be less than 0.1\%. 

The low-lying states of $^7$He are expected, based on simple considerations and supported by sophisticated structure model calculations (Table S1~\cite{supp}), to arise from a $p$-shell neutron coupled to the g.s. or  $2^{+}_1$ states of $^{6}$He. 
Given that the $^{8}$He g.s. is dominated by a closed neutron $(p_{3/2})^4$ configuration \cite{Skaza06}, the most strongly populated level here is, as observed, the $^{7}$He($3/2^{-}_1$) g.s.. The predicted $1/2^{-}$ first excited state could, in principle, be populated through neutron removal from the expected small $(p_{3/2})^{2}(p_{1/2})^2 $ configuration, but its strength should be weak, consistent with the lack of any clear signature here for such a state in the $^{6}$He+$n$ channel. 

As such, the structure observed here at higher energy [$E_x=2.68(4)$~MeV] in the $^{4}$He+3$n$ channel likely arises from either or both of the predicted $3/2^{-}_2$ and $5/2^{-}_1$ levels (Table S1~\cite{supp}). The first report of such a structure \cite{Kor99} preferred a $5/2^{-}_1$ assignment, based on the absence of $3/2^{-}_2$ built on $^{6}$He$^*(2^{+})$ in RRGM calculations \cite{Wurzer98}. Subsequent modeling, including those in Table S1, indicate that the exclusion of $3/2^{-}_2$ was premature. 
In the current measurement, the momentum distribution for this structure is identical to that for the $^{7}$He g.s. (Fig.~\ref{fig:3n} inset), indicating its production via $p$-shell neutron knockout.  
The structure of $5/2^{-}_1$ (Table S1 \cite{supp}) implies that it can only be produced in a two-step process requiring excitation of the $^6$He core to its 2$^+_1$ state, while $3/2^{-}_2$ can be produced directly via the small $^6$He$(0^+) \otimes p_{3/2}$ admixture.  In this context, we note that, excluding the CSM calculations \cite{Myo07}, the ratio of the predicted spectroscopic strengths for the $^6$He$(0^+)\otimes p_{3/2}$ configuration of the $3/2^{-}_{1,2}$ states is similar to the ratio of their measured efficiency-corrected yields ($\sim$10).  As such, we conclude that the structure observed here arises predominately from $3/2^{-}_2$ with a small contribution from $5/2^{-}_1$ possible.

To further support these arguments, Table S1 \cite{supp} includes new GSM calculations \cite{Michel09, Michel02, Xie23,  Xie23b} performed for the present study. These calculations incorporate a $^4$He core and include all partial waves up to $l$=3 for the three valence neutrons. The interactions are adopted from Ref.~\cite{Jaganathen17}, which employs a Woods-Saxon potential with a spin-orbit term for the core-$n$ one-body interaction and the Furutani-Horiuchi-Tamagaki (FHT) $n$--$n$ interaction \cite{Furutani78, Furutani79} . These calculations predict a $5/2^-_1$ level with $E_x$=$2.73$~MeV and $3/2^{-}_2$ with $E_x$=$3.37$~MeV, both resonances being rather broad ($\Gamma$$\sim$2--2.5~MeV) and built predominately on the $^6$He$(2^+)$$\otimes$$p_{1/2}$ configuration, with the latter also exhibiting a small $^6$He$(0^+)$$\otimes$$p_{3/2}$ admixture.
 
\begin{figure}[htbp]
\includegraphics[width=0.7\linewidth]{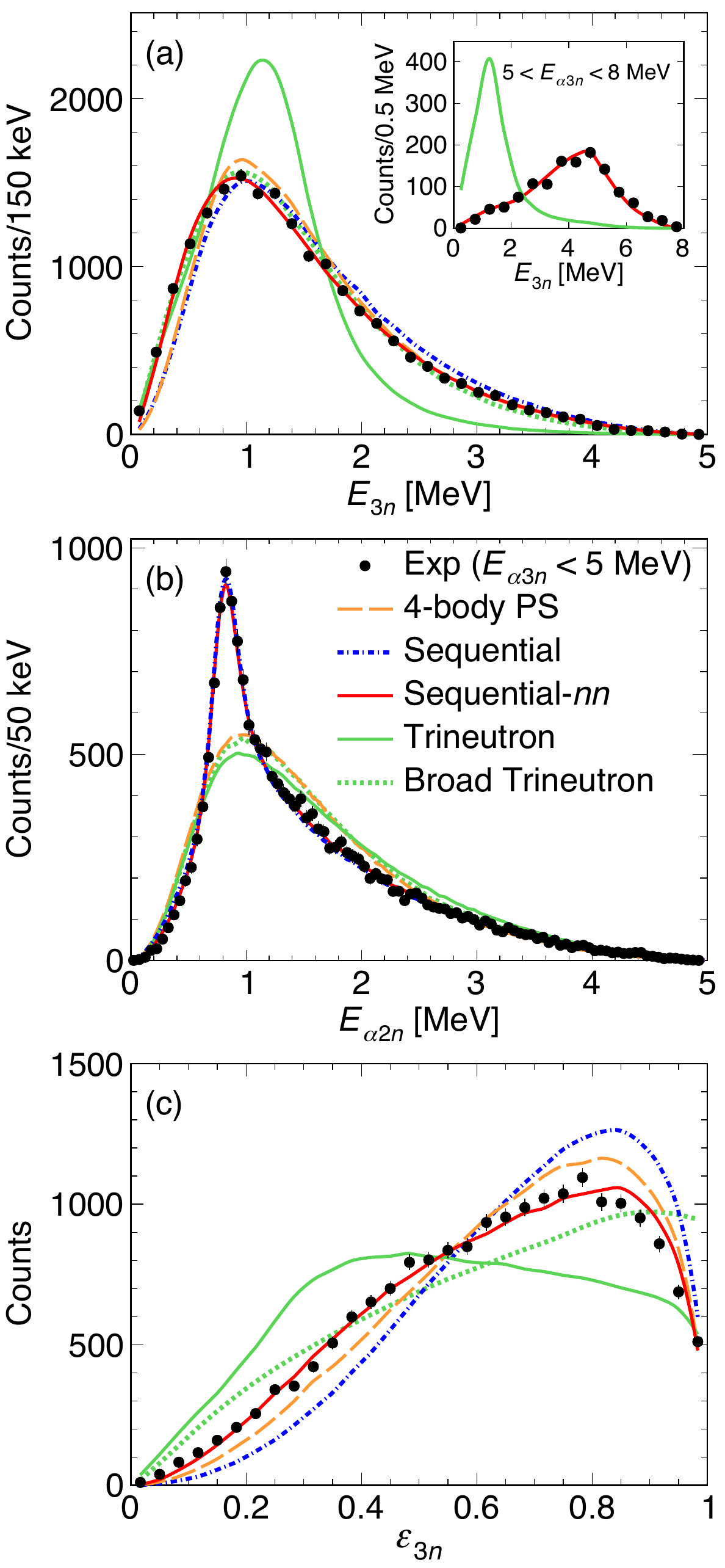}
\caption{Relative-energy spectra for $^{4}$He+3$n$ events with $E_{\alpha3n}<5$~MeV:
 (a) $3n$ energy $E_{3n}$;
 (b) energy, $E_{\alpha2n}$, of the $^{4}$He+2$n$ subsystem for the $^{4}$He+3$n$ events;
 (c) normalized $3n$ energy $\varepsilon_{3n}=E_{3n}/E_{\alpha3n}$.
 The lines correspond to simulations of different decay scenarios (see text). The inset of panel (a) shows the $E_{3n}$ spectrum for events with $5<E_{\alpha3n}<8$~MeV. The vertical error bars are statistical only.}
\label{fig:3ncorrelation}
\end{figure}

\header{The three-neutron system} 
Turning to the $3n$ system itself, Fig.~\ref{fig:3ncorrelation}a shows for the first time an $E_{3n}$ spectrum reconstructed from the invariant mass of three neutrons for $^{4}$He+3$n$ events with $E_{\alpha3n}$$<$5~MeV.  Prior to detailed discussion of possible $3n$ correlations, it is worth examining the relative energy of the $^{4}$He+2$n$ subsystem reconstructed from the $^{4}$He+3$n$ events (Fig.~\ref{fig:3ncorrelation}b) --- i.e., including all three possible $^{4}$He+$2n$ combinations. 
The narrow peak at $E_{\alpha2n}$$\sim$0.8~MeV, characteristic of $^{6}$He$^*(2^{+}_1)$, is clearly observed, but superimposed on a much broader component. This suggests that the $3n$ decay is sequential, as in this process one of the three $^{4}$He+$2n$ combinations corresponds to the feeding of $^6$He$^*(2^+_1)$ while the other two, in which one neutron arises from the decay of $^{7}$He$^{*}$ to $^{6}$He$^{*}(2^{+}_1)$, result in the broader component. 

In the following, we compare the relative-energy spectra to four representative decay scenarios using complete GEANT4 simulations described above, each using as input the experimental $E_{\alpha3n}$ spectrum shown in Fig.~\ref{fig:3n}.
All the simulations are normalized according to the total number of events.

\begin{itemize}
\item ``4-body PS'' (orange dashed line): where the decay of $^7$He$^*$ into $^{4}$He+3$n$ is governed by 4-body phase space (PS) without any $^{4}$He-$n$ or $n$-$n$ final state interactions (FSI).

\item ``Trineutron'': where the decay is initially 2-body into $^{4}$He+$^3n$, using the trineutron resonance parameters predicted by NCGSM \cite{Li19} (green solid line, $E_{3n}=1.3$~MeV and $\Gamma_{3n}=0.9$~MeV).The $^3n$ then undergoes 3-body PS decay without any $^{4}$He-$n$ or $n$-$n$ FSI. For comparison, the ``Broad Trineutron'' case is also considered (green dashed line), with trineutron resonance parameters obtained from a fit to the experimental $E_{3n}$ spectrum.

\item ``Sequential'' (blue dash-dotted line): where the first step is a 2-body decay into $^6$He$^*(2^+_1)$+$n$, using the $^6$He$^*(2^+_1)$ resonance parameters obtained above. The subsequent decay of $^6$He$^*(2^+_1)$ into $^{4}$He+2$n$ is modeled as a pure 3-body PS process, without any $^{4}$He-$n$ or $n$-$n$ FSI.

\item ``Sequential-$nn$'' (red solid line): the ``Sequential'' scenario with the addition of the effects of the $n$-$n$ interaction in the decay of $^6$He$^*(2^+_1)$, included as the emission of the neutron pair in an $s$-wave virtual state \cite{Esbensen97} ($a_s$~=~-18.7~fm \cite{Trotter06}). No $^4$He-$n$ FSI  is included. Such a scenario is consistent with the strong $n$-$n$ correlations exhibited by the $^{4}$He+2$n$ events (Fig.~S3 \cite{supp}).
\end{itemize}

The $E_{3n}$ spectrum (Fig.~\ref{fig:3ncorrelation}a), which exhibits a broad structure peaking at around 1~MeV, is well described by all the scenarios except ``Trineutron'', which, as expected, is much narrower, reflecting the relatively narrow intrinsic width of the predicted resonance \cite{Li19}. This difference is further amplified when events from the high-energy tail of $^7$He$^*$ ($5<E_{\alpha3n}<8$~MeV) are considered, as shown in the inset of Fig.~\ref{fig:3ncorrelation}a. We also note that other trineutron resonance parameters ($E_{3n}=0.35$~MeV and $\Gamma_{3n}=0.7$~MeV; $E_{3n}=0.5$~MeV and $\Gamma_{3n}=1$~MeV) predicted by NCSM  \cite{Mazur24} yield similarly narrow $E_{3n}$ distributions, with peaks even closer to the threshold.

Clearly, the $E_{\alpha2n}$ spectrum of the $^{4}$He+2$n$ subsystem (Fig.~\ref{fig:3ncorrelation}b) is only consistent with the two sequential decay scenarios. Direct emission of the three neutrons, either via 4-body phase space or as a trineutron resonance, cannot reproduce the prominent peak at $E_{\alpha2n}$$\approx$0.8~MeV, which arises from decay via $^6$He$^*(2^+_1)$. 
For ``Trineutron'', while better agreement with the $E_{3n}$ spectrum can be achieved by broadening the assumed trineutron resonance in the simulation (``Broad Trineutron'' shown as the green dashed line in Fig.~\ref{fig:3ncorrelation}), it remains impossible to reproduce the very peaked distribution that characterizes the $E_{\alpha2n}$ spectrum nor can the $\varepsilon_{3n}$ (see below) spectrum be described. We have also verified that the sequential decay via the $^5$He g.s. resonance does not play a measurable role and, indeed, no evidence is seen for its formation (Fig.~S3 \cite{supp}).
The $3n$ decay of the $^7$He$^*$ structure observed here is thus predominantly sequential via $^6$He$^*(2^+_1)$. 

Regarding the two sequential scenarios, the shape of the $E_{3n}$ spectrum (Fig.~\ref{fig:3ncorrelation}a) is better reproduced by including the $n$--$n$ interaction (``Sequential-$nn$'') which shifts the distribution slightly toward lower energies. 

In Fig.~\ref{fig:3ncorrelation}c, we display the normalized $3n$ energy $\varepsilon_{3n}=E_{3n}/E_{\alpha3n}$, defined as the ratio of the $3n$ energy to the total energy. 
This variable rescales the $3n$ energy spectrum to a normalized range from 0 to 1 and, as shown, is more sensitive to the different scenarios.  Again, the sequential scenario incorporating the $n$--$n$ interaction (``Sequential-$nn$'') in the $^6$He$^*(2^+_1)$ decay provides the best description.

Finally, we performed a $\chi^2$ fit to the $E_{\alpha2n}$ spectrum using a combination of ``Sequential-$nn$'' and ``Trineutron'' in order to set an upper limit for the latter, yielding 0.8\% at the $1\sigma$ level.
If we consider a broader trineutron resonance (``Broad Trineutron'': $E_{3n}=3$~MeV and $\Gamma_{3n}=5$~MeV) that matches more closely the $E_{3n}$ spectrum of Fig.~\ref{fig:3ncorrelation}a, an upper limit of 1.6\% can be set. Note that a further increase in the resonance width, such as the broad ${3n}$ structure reported recently in a measurement of the $^3$H$(t,^3$He$)3n$ reaction ($\Gamma_{3n}\approx15$~MeV) \cite{Miki24}, leads to an even smaller upper limit.

It is therefore concluded that the $3n$ emission from the $^7$He$^*$ structure observed here is consistent with a purely sequential process via $^6$He$^*(2^+_1)$.  In addition, no evidence is found for a narrow, relatively low-lying trineutron resonance such as that predicted  \cite{Gandolfi17, Li19, Mazur24}.  While significant 2-body neutron correlations are observed in the decay of $^6$He$^*(2^+_1)$, similar to those observed in the dissociation of $^6$He \cite{Aumann99} and in the $^3$H$(\pi^-,\gamma)3n$ reaction \cite{Miller80}, no evidence is found for any significant $3n$ correlations.

\header{Conclusions} 
In summary, $3n$ emission from $^7$He has been directly measured for the first time. 
A prominent resonance-like structure lying $2.08(4)$~MeV above the $^4$He$+3n$ threshold with a width of $3.9(2)$~MeV was observed.  Based on the momentum distribution of the knocked-out neutron and comparison with state-of-the-art structure model calculations, it was attributed to arising predominately from the predicted $3/2_2^-$ excited state of $^7$He with a possible small admixture from $5/2_1^-$.
The $3n$ invariant-mass spectrum and the underlying neutron correlations were investigated.  A broad structure peaking at around 1~MeV was observed in the $E_{3n}$ spectrum and, combined with the $E_{\alpha2n}$ spectrum and complete simulations incorporating $n$--$n$ correlations, was very well reproduced by the sequential decay of $^7$He$^*$ via $^6$He$^*(2_1^+)$.
No evidence was found for any significant three-neutron correlations, including any signature of a trineutron resonance. The high-quality three-neutron-emission data reported here could provide a benchmark for theoretical descriptions of multi-neutron correlations and decay mechanisms of unbound neutron-rich systems.

The present work underlines that direct multineutron detection and the subsequent neutron correlation analyses hold much promise for understanding the structure of multineutron unbound states and the nature of multineutron systems. Such measurements are being extended to other multineutron emitters, such as $4n$ unbound $^7$H.

\acknowledgments {We would like to thank the RIBF accelerator staff for the primary beam delivery and the BigRIPS team for their efforts in preparing the secondary beams. Instructive discussions with Prof. Tetsuo Noro (Kyushu University) are gratefully acknowledged. S. H. and Z. Y. acknowledge the support from the National Key R$\&$D Program of China (Grants No. 2023YFE0101500 and 2022YFA1605100), the National Natural Science Foundation of China (Grant No. 12275006). The LPC Caen participants acknowledges partial support from the French-Japanese LIA-International Associated Laboratory for Nuclear Structure Problems as well as the French ANR14-CE33-0022-02 EXPAND. P. L. acknowledges the support from the National Natural Science Foundation of China (Grant No. 12405147). T. U. acknowledges support from the Grants-in-Aid of the Japan Society for the Promotion of Science (Grant No. JP21H04975).This project was supported by Deutsche Forschungsgemeinschaft (DFG, German Research Foundation), Project-ID 279384907---SFB 1245, the GSI-TU Darmstadt cooperation agreement, the JSPS A3 Foresight Program, JSPS KAKENHI Grant No. JP24H00006 and JP18H05404, the Institute for Basic Science (IBS-R031-D1), BMBF (05P24PK1), the State Key Laboratory of Nuclear Physics and Technology, Peking University (No. NPT2024ZX01, No. NPT2025ZX02). S. P. acknowledges support by the UK STFC under contract numbers ST/P003885/1 and ST/L005727/1 and the University of York Pump Priming Fund. S. H. also acknowledges the support from the International Program Associate of RIKEN and the National Research Foundation of Korea (NRF) and the Ministry of Science and ICT (RS-2024-00436392). }

\end{document}